\documentclass[pra,aps,10pt,twocolumn,amsmath,amssymb]{revtex4-1}

\usepackage[english]{babel}
\usepackage{graphicx}% Include figure files
\usepackage{bm}% bold math
\usepackage{color}
\usepackage{slashed}%slashed D, k, A, etc.
\usepackage{latexsym}
\usepackage{appendix}

\newcommand{\dd}{d}
\newcommand{\ii}{i}
\newcommand{\ee}{e}

\newcommand{\me}{m_\mathrm{e}} %electron mass
\newcommand{\op}[1]{\hat{#1}}

\begin{document}

\title{Photoelectron combs in ionization: Influence of rescattering and nondipole effects}
\author{J. Z. Kami\'nski} \email{Jerzy.Kaminski@fuw.edu.pl}
\author{K. Krajewska} \email{Katarzyna.Krajewska@fuw.edu.pl}

\affiliation{
Institute of Theoretical Physics, Faculty of Physics, University of Warsaw, Pasteura 5, 02-093 Warsaw, Poland }
\date{\today}

\begin{abstract}
Ionization by a sequence of extreme ultraviolet pulses is investigated based on the rigorous numerical solution of the time-dependent Schr\"odinger equation, when
the driving laser field is treated exactly. This goes beyond the typically used first-order nondipole approximation and reveals the effects of 
radiation pressure to its full extent. Specifically, we observe the comb structures in both the momentum and the energy distributions 
of photoelectrons. The comb peaks are shifted, however, depending on the emission angle of electrons. While similar effect is observed already 
in the first-order nondipole approximation, with increasing the laser field strength the discrepancy with our exact results becomes more pronounced.
Also, we observe the additional substructure of the comb peaks arising in the angle-integrated energy distributions of photoelectrons. 
Finally, as our numerical calculations account for the atomic potential in the 
entire interaction region, we observe the loss of coherence of comb structures with increasing the number of laser pulses, that we attribute to 
rescattering.
\end{abstract}

\maketitle

\section{Introduction}
\label{sec:introduction}

Since the first demonstration of attosecond extreme ultraviolet (XUV) pulses~\cite{Paul2001,Hentschel}, they have become a prominent diagnostic tool to study ultrafast electron
dynamics in matter~\cite{Agostini,Krausz,LHuillier}. For instance, a pump-probe scenario combining trains of attosecond laser pulses with infrared (IR) laser fields is now 
the standard method of attosecond photoelectron interferometry~\cite{Remetter}. Interestingly, the method has been
extended recently into the nondipole regime~\cite{Worner2024}. As, in general, it requires good experimental stability and
precise control, it is also highly valuable to develop more refined theoretical models to precisely capture the underlying physics.

Laser-matter interactions are typically described in the dipole approximation, in which the laser field is treated as a time-dependent electric component.
Such treatment fails, however, for high-frequency or low-frequency but large intensity laser fields~\cite{Reiss2014}. Under such circumstances, nondipole effects become important 
as the field imparts momentum on the electron and the magnetic field component plays a role (for recent reviews on nondipole effects, see Refs.~\cite{Peng,Keller}). 
These factors are disregarded by the dipole 
approximation. A step forward was made in theory by relying on the first-order $1/c$ expansion of the electron Hamiltonian in an external laser 
field and numerically solving the time-dependent Schr\"odinger equation (TDSE). In this framework, aspects such as the forward-backward asymmetry of 
photoelectron momentum distributions~\cite{Forre2006,Lein2018,Hartung,Burgdorfer,LeiGeng}, momentum sharing between the electron and its parent 
ion~\cite{Chelkowski,HartungMAG,Lin}, or nondipole contributions to photoionization time delays~\cite{Pi2024} were investigated. 
However, few works already have demonstrated deviations of the first-order nondipole approximation 
from the exact solution of the TDSE where the laser-matter interaction is treated rigorously~\cite{Suster,SusterOPT,KK2025,KK2025OPT}.

In order to provide reliable theoretical tools for future attosecond metrology, in the current paper we consider ionization of hydrogen atom by 
a train of XUV pulses. In contrast to the previous studies~\cite{Collins,Nisoli,Peng2014,Ciappina}, we rigorously account for the XUV field, i.e., for its 
space and time dependence. This goes beyond the current status of XUV-pump IR-probe experiments. Specifically, in Ref.~\cite{Worner2024} 
it was enough to describe the interaction with external fields in the first-order nondipole approximation; while the interaction with XUV pulses
was treated in the electric quadrupole approximation, the magnetic field effects have been exclusively accounted for through the magnetic dipole
coupling to a stronger IR pulse.
Still, with an ongoing progress in laser technologies, we find it interesting to shine some
light on effects that can be potentially uncovered by XUV-pump XUV-probe experiments. Note that some experimental progress in this area has been
already made~\cite{Okino,Tzallas}.

It is known that photoelectron comb structures arise when atomic or molecular targets are ionized by a controlled sequence of laser pulses. 
For instance, for a low-frequency 
and high-intensity train of pulses, this was theoretically demonstrated in the relativistic domain using the strong-field approximation 
(SFA)~\cite{CombsPLA}. Note that in this approximation the electron interaction with the residual ion is neglected. At the same time, the 
Fraunhoffer-type formula was derived there, revealing that the comb structures originate from interference of ionization probability amplitudes 
from individual pulses comprising the train; in other words, from interpulse interference. Since numerical and analytical results were based on the SFA, the perfect coherent
enhancement of ionization distributions was predicted. This is in contrast to Ref.~\cite{Eikonal}, where the generalized eikonal approximation was used 
to treat ionization by an IR train of pulses but in the nonrelativistic regime. In this approach, rescattering phenomena were accounted for, 
which in turn affected the coherence of comb structures. The IR field, however, was treated within the dipole approximation.
The purpose of the current paper is to study how this picture will change if both rescattering and nondipole effects are treated simultaneously.
To this end, we will solve numerically the TDSE with the exact minimal coupling Hamiltonian, using the Suzuki-Trotter 
scheme with the split-step Fourier method~\cite{Suster}. As such calculations are computationally demanding, we shall study a two-dimensional system. 
In our configuration we will be able to control the leakage of the electron wave function from the interaction region, limiting it to a small value 
($10^{-8}$ for the total probability). Hence, providing precise theoretical predictions of photoelectron momentum and energy distributions.

As a background for our numerical investigations, we recall in Sec.~\ref{sec::sfa} the foundations of the quasi-relativistic strong-field approximation 
(QRSFA)~\cite{Ehlotzky,EhlotzkyReview,KK2015}. Note that it is related to our TDSE approach in the sense that both treat the external electromagnetic field exactly.
This also distinguished QRSFA from other SFA-based approaches as, for instance, those published in Refs.~\cite{He,Boning2019,Hasibovic}.
Moreover, QRSFA allows one to interpret nondipole effects as arising from either retardation or electron recoil. Therefore, before we present our TDSE
results, we adapt the QRSFA method to describe ionization by a train of identical pulses. This will result in the development of the Fraunhoffer-like
formula and predicting the appearance of ionization comb structures under conditions considered in our paper. 

Our TDSE results are demonstrated and discussed in detail in Secs.~\ref{nodelay},~\ref{timedelay}, and~\ref{comparison}. We start by considering 
ionization by an XUV train of pulses without time delay between them. We observe well-resolved photoelectron combs in the momentum-angular
distributions of photoelectrons. Notably, they are characterized by a series of tilted fringes, with an inclination depending on the electron
energy and its emission angle. Energy distributions, on the other hand, are presented for the given photoelectron emission angle, exhibiting a clear 
forward-backward asymmetry. In contrast to the QRSFA prediction, the observed comb structures do not increase in magnitude coherently with the number 
of pulses in the train. This is attributed to rescattering effects which are absent in QRSFA. As shown in Sec.~\ref{timedelay}, the energy
spacing of the combs decreases as the time delay between the subsequent pulses from the train increases. Still, for the given parameters, the comb
peaks are very well-resolved. In light of these results, a valid question arises: How do they compare to the first-order nondipole
or to the dipole approximate results? This is discussed in Sec.~\ref{comparison}. Finally, we summarize our investigations in Sec.~\ref{conclusions}.

In our TDSE calculations, we use the atomic units of momentum $p_0=\alpha\me c$, energy $E_0=\alpha^2\me c^2$, length $a_0=\hbar/p_0$, time 
$t_0=\hbar/E_0$, and the electric field strength $\mathcal{E}_0=\alpha^3\me^2 c^3/(|e|\hbar)$, where $\me$ and $e=-|e|$ are the electron rest mass 
and charge, whereas $\alpha$ is the fine-structure constant. In analytical formulas, on the other hand, we set $\hbar=1$ while keeping explicitly 
the remaining fundamental constants.

\section{QRSFA prediction of comb structures in ionization}
\label{sec::sfa}

We start with the derivation of the Fraunhoffer formula for ionization by a sequence of identical laser pulses, using the QRSFA~\cite{KK2015}.
This will set the grounds for our further investigations in Sections~\ref{nodelay} and~\ref{timedelay}, which will be based on  
numerical integration of the two-dimensional time-dependent Schr\"odinger equation. 

\subsection{General SFA approach}

Consider a two-dimensional single-electron system in a laser field, whose Hamiltonian is given as
\begin{equation}
H(t)=H_0+H_I(t).
\label{sfa1}
\end{equation}
Here, $H_0$ is the static atomic Hamiltonian, whereas $H_I(t)$ describes the laser-matter coupling. Generally, one assumes that the interaction
Hamiltonian $H_I(t)$ vanishes asymptotically in the remote past and in the far future, i.e., $H_I(t)\rightarrow 0$ for $t\rightarrow\pm\infty$. 
%As we shall focus on ionization by a laser field lasting for a finite time $T_p$, this assumption is naturally satisfied. 
We define the atomic bound state $|\psi_0\rangle$ 
of energy $E_0$, meaning that $H_0|\psi_0\rangle=E_0|\psi_0\rangle$. Hence, the transition probability amplitude from the bound state $|\psi_0\rangle$ 
to the scattering state $|\psi_{\bm p}(t)\rangle$, the latter specified by the electron asymptotic momentum $\bm{p}$, is
\begin{equation}
\mathcal{A}(\bm{p})=-\ii\int \dd t \langle\psi_{\bm{p}}(t)|H_I(t)|\psi_0\rangle \ee^{-\ii E_0t}.
\label{sfa2}
\end{equation}
While $|\psi_{\bm{p}}(t)\rangle$ is in principle the exact scattering solution of the full Hamiltonian~\eqref{sfa1}, for $t\rightarrow\infty$
it goes to the specific scattering state of the atomic Hamiltonian $H_0$ with the incoming spherical wave boundary condition. As this state is 
difficult to determine, one proceeds with approximations. Specifically, in the SFA, the exact scattering state
$|\psi_{\bm p}(t)\rangle$ is replaced by the Volkov state, $|\psi_{\bm p}^{(0)}(t)\rangle$, which amounts to
neglecting the interaction of the ionized electron with the residual ion~\cite{Volkov}. In other words, it solves the equation
\begin{equation}
\ii \partial_t |\psi^{(0)}_{\bm{p}}(t)\rangle=[H_{\mathrm{kin}}+H_I(t)]|\psi^{(0)}_{\bm{p}}(t)\rangle,
\label{sfa4}
\end{equation}
where $H_{\mathrm{kin}}$ is the kinetic energy portion of $H_0$. This indicates that the SFA is equivalent to the lowest order Born approximation 
with respect to the binding potential. It should be justified, therefore, when the photoelectron kinetic energy $E_{\bm p}$ 
is large compared to its binding energy $E_0$ or when the influence of the laser field dominates the atomic interaction.  Under those circumstances,
the probability amplitude~\eqref{sfa2} becomes
\begin{equation}
\mathcal{A}_{\mathrm{SFA}}(\bm{p})=-\ii\int \dd t \langle\psi^{(0)}_{\bm{p}}(t)|H_I(t)|\psi_0\rangle \ee^{-\ii E_0t},
\label{sfa3}
\end{equation}
which is the starting point of our further derivations.

\subsection{QRSFA}

Below, we shall use the SFA to predict the structure of photoelectron momentum distributions when ionization is driven by a finite train of identical 
pulses. Our current derivation will be based on the quasi-relativistic version of the SFA, that was developed in Ref.~\cite{KK2015}. For this purpose, we 
shall use the shorthand relativistic notation: $k=(k^0,{\bm k})=k^0(1,{\bm n})=\frac{\omega}{c}(1,{\bm n})$, 
$x=(x^0,{\bm x})=(ct,{\bm x})$, and $p=(p^0,{\bm p})=(\frac{E_{\bm p}}{c},{\bm p})$, where the spatial-like coordinates represent vectors
in the two-dimensional space. Keeping this in mind, Eq.~\eqref{sfa3} can be represented as
\begin{align}
\mathcal{A}_{\mathrm{SFA}}(\bm{p})=-\ii\int\frac{\dd^2 q}{(2\pi)^2}\int \dd^3 x \bigl[&\psi^{(0)}_{\bm{p}}(x)\bigr]^{*}\frac{1}{c}H_I(x) \nonumber \\
\times &\tilde{\psi}_0(\bm{q}) \ee^{-\ii q\cdot x},
\label{sfa5}
\end{align}
where 
\begin{equation}
\psi_0(\bm{x})=\int\frac{\dd^2 q}{(2\pi)^2}\,\tilde{\psi}_0(\bm{q})\ee^{\ii \bm{q}\cdot\bm{x}}
\label{sfa6}
\end{equation}
and $q=(q^0,\bm{q})$ with $q^0=E_0/c$. Note that we introduce also the relativistic-like scalar product, $q\cdot x=q^0x^0-{\bm q}\cdot {\bm x}$; the convention that will be used below 
for other scalar products as well. 

In the QRSFA, the space-dependence of the laser field is taken into account exactly. This is evident in the interaction 
Hamiltonian
\begin{equation}
\frac{1}{c}H_I(x)=\ii\frac{e\bm{A}(k\cdot x)\cdot\bm{\nabla}}{m_{\mathrm{e}}c}+\frac{e^2\bm{A}^2(k\cdot x)}{2m_{\mathrm{e}}c},
\label{qrsfa1}
\end{equation}
as the vector potential phase equals $k\cdot x=k^0(x^0-{\bm n}\cdot {\bm x})=\omega(t-{\bm n}\cdot{\bm x}/c)=\omega t_{\rm r}$.
Here, the retarded time $t_{\rm r}$ is defined implicitly. The same concerns the Volkov state, which is approximated as~\cite{Ehlotzky,EhlotzkyReview,KK2015}
\begin{align}
\psi^{(0)}_{\bm{p}}({\bm x},t)=\frac{1}{\sqrt{V}}\exp\Bigl[&-\ii p\cdot x +\ii\int_0^{k\cdot x}\dd\phi'\Bigl(\frac{e{\bm A}(\phi')\cdot {\bm p} }
{m_{\mathrm{e}}ck^0-\bm{p}\cdot\bm{k}}\nonumber\\
&-\frac{e^2{\bm A}^2(\phi')}{2(m_{\mathrm{e}}ck^0-\bm{p}\cdot\bm{k})}\Bigr)\Bigr].
\label{qrsfa2}
\end{align}
While the radiation pressure effects and high laser field intensity effects are accounted for in Eq.~\eqref{qrsfa2}, it does neglect the relativistic 
corrections such as the mass correction or the spin-orbit interaction, which are of the order of $1/c^2$. Therefore, the electron kinetic energy, which  is present there through the term
$p\cdot x$, equals $E_{\bm p}={\bm p}^2/2m_{\rm e}$. Also, $V$ in Eq.~\eqref{qrsfa2} denotes the quantization volume. With the above formulas and
following Ref.~\cite{KK2015}, we reduce Eq.~\eqref{sfa5} to the one-dimensional integral over the laser field phase $\phi=k\cdot x$,
\begin{widetext}
\begin{align}
{\cal A}_{\rm QRSFA}({\bm p})&=\frac{\ii}{\sqrt{V}}\frac{1}{k^0}\tilde{\psi}_0({\bm p}+(p^0-q^0){\bm n})\nonumber\\
&\times\int\dd\phi
\left(\frac{e\bm{A}(\phi)\cdot\bm{p}}{m_{\mathrm{e}}c}-\frac{e^2\bm{A}^2(\phi)}{2m_{\mathrm{e}}c}\right)\exp\left[\ii\int_0^{\phi}\dd\phi'
\Bigl(\frac{p^0-q^0}{k^0}-\frac{e{\bm A}(\phi')\cdot {\bm p} }{m_{\mathrm{e}}ck^0-\bm{p}\cdot\bm{k}}
+\frac{e^2{\bm A}^2(\phi')}{2(m_{\mathrm{e}}ck^0-\bm{p}\cdot\bm{k})}\Bigr)\right].
\label{qrsfa-new}
\end{align}
\end{widetext}
Note that the integration limits in the integral over $\phi$ are specified below, where we adapt Eq.~\eqref{qrsfa-new}
to describe ionization by a finite sequence of laser pulses.

\subsection{Fraunhoffer formula}

Let us focus now on ionization by a train of $N_{\rm rep}$ identical pulses, being the main focus of our paper. As it will be introduced in
Sections~\ref{nodelay} and~\ref{timedelay}, the corresponding vector potential equals ${\bm A}(\phi)=A_0{\bm{\varepsilon}}f(\phi)$, where the shape function $f(\phi)$ is nonzero 
only for phases $\phi$ within the interval $[0,2\pi N_{\rm rep}]$. This condition sets the integration limits in Eq.~\eqref{qrsfa-new}.
In addition, for $\phi\in [0,2\pi]$ it holds,
\begin{equation}
f(\phi)=f(\phi+2\pi (\ell-1)),
\label{abbrev4}
\end{equation}
where $\ell=1,2,...,N_{\rm rep}$ and $f(0)=f(2\pi N_{\rm rep})=0$. To shorten the notation, we define
\begin{align}
F({\bm p})=\frac{p^0-q^0}{k^0}&-\frac{eA_0 ({\bm{\varepsilon}}\cdot {\bm p})}{m_{\mathrm{e}}ck^0-\bm{p}\cdot\bm{k}}\langle f\rangle\nonumber\\
&+\frac{e^2A_0^2}{2(m_{\mathrm{e}}ck^0-\bm{p}\cdot\bm{k})}\langle f^2\rangle,
\label{abbrev5}
\end{align}
with the phase-averaged values of the shape function,
\begin{equation}
\langle f^j\rangle=\frac{1}{2\pi}\int_0^{2\pi}\dd\,\phi f^j(\phi), \quad {\rm for}\; j=1,2.
\label{abbrev3}
\end{equation}
In addition, we introduce
\begin{align}
M(\phi;{\bm p})&=\frac{\ii}{\sqrt{V}}\tilde{\psi}_0({\bm p}+(p^0-q^0){\bm n})\nonumber\\
&\times\left[\frac{eA_0}{m_{\mathrm{e}}c}({\bm{\varepsilon}}\cdot\bm{p})f(\phi)
-\frac{e^2A_0^2}{2m_{\mathrm{e}}c}f^2(\phi)\right],\label{abbrev1}\\
G_{\rm osc}(\phi;{\bm p})&=-\frac{eA_0 ({\bm{\varepsilon}}\cdot {\bm p})}{m_{\mathrm{e}}ck^0-\bm{p}\cdot\bm{k}}[f(\phi)-\langle f\rangle]\nonumber\\
&+\frac{e^2A_0^2}{2(m_{\mathrm{e}}ck^0-\bm{p}\cdot\bm{k})}[f^2(\phi)-\langle f^2\rangle],
\label{abbrev2}
\end{align}
which satisfy the same periodicity condition as $f(\phi)$ [Eq. \eqref{abbrev4}]. This enables us to express the ionization probability 
amplitude~\eqref{qrsfa-new},
\begin{equation}
{\cal A}_{\rm QRSFA}({\bm p})=\frac{1}{k^0}\int_0^{2\pi N_{\rm rep}}\dd\phi M(\phi)\ee^{\ii F({\bm p})\phi+\ii G_{\rm osc}({\bm p};\phi)},
\label{abbrev6}
\end{equation}
as a coherent superposition of probability amplitudes resulting from individual pulses. Starting with
\begin{equation}
{\cal A}_{\rm QRSFA}({\bm p})=\sum_{\ell=1}^{N_{\rm rep}}{\cal A}_{\rm QRSFA}^{(\ell)}({\bm p}),
\label{abbrev7}
\end{equation}
where
\begin{equation}
{\cal A}_{\rm QRSFA}^{(\ell)}({\bm p})=\frac{1}{k^0}\int_{2\pi(\ell-1)}^{2\pi\ell}\dd\phi M(\phi)\ee^{\ii F({\bm p})\phi
+\ii G_{\rm osc}({\bm p};\phi)},
\label{abbrev8}
\end{equation}
we obtain 
\begin{equation}
{\cal A}_{\rm QRSFA}({\bm p})=\sum_{\ell=1}^{N_{\rm rep}}\ee^{2\pi\ii (\ell-1)F({\bm p})}{\cal A}_{\rm QRSFA}^{(1)}({\bm p}).
\label{abbrev9}
\end{equation}
Here, ${\cal A}_{\rm QRSFA}^{(1)}({\bm p})$ is the probability amplitude of ionization by a single pulse from the train. Summing up the remaining
expression, we end up with the following Fraunhoffer formula,
\begin{align}
{\cal A}_{\rm QRSFA}({\bm p})&=\ee^{\ii\pi N_{\rm rep}F({\bm p})}\nonumber\\
&\times\frac{\sin[\pi N_{\rm rep}F({\bm p})]}{\sin[\pi F({\bm p})]}
{\cal A}_{\rm QRSFA}^{(1)}({\bm p}).
\label{abbrev10}
\end{align}
As one can see, except of the overall phase factor, the probability amplitude of ionization by $N_{\rm rep}$ identical laser pulses is proportional
to the probability amplitude of ionization by a single pulse. In addition, it is modulated by the interference term containing the quotient of two 
sine functions. The latter leads to the coherent $N_{\rm rep}^2$-like enhancement of the probability distributions. In other words, 
we expect that at certain photoelectron momenta ${\bm p}_n$ such that $\pi F({\bm p}_n)=n\pi$, $n\in\mathbb{Z}$,
the probability distribution of ionization by $N_{\rm rep}$ identical pulses will be $N_{\rm rep}^2$ times enhanced as compared to the one by the single pulse. Also, between every two
such momenta, ${\bm p}_n$ and ${\bm p}_{n+1}$, there will be $(N_{\rm rep}-1)$ zeros in the distribution. This suggests the appearance of the major peaks at ${\bm p}_n$
and $(N_{\rm rep}-2)$ secondary peaks in between them. Note that this picture holds for as long as the Born approximation is applicable. Therefore,
in the following sections we shall confront it with the results of purely numerical analysis, which accounts for the interaction of the ionized electron and its 
parent ion. 

In closing this Section, we note that although Eq.~\eqref{abbrev10} has been derived for a two-dimensional system, the same formula will hold in 
three dimensions as well. The only difference is that the two-dimensional vectors must then be replaced by their three-dimensional counterparts.

\section{Photoelectron combs generated by a pulse train with no time delay}
\label{nodelay}
\begin{figure}[t]
\includegraphics[width=6.5cm]{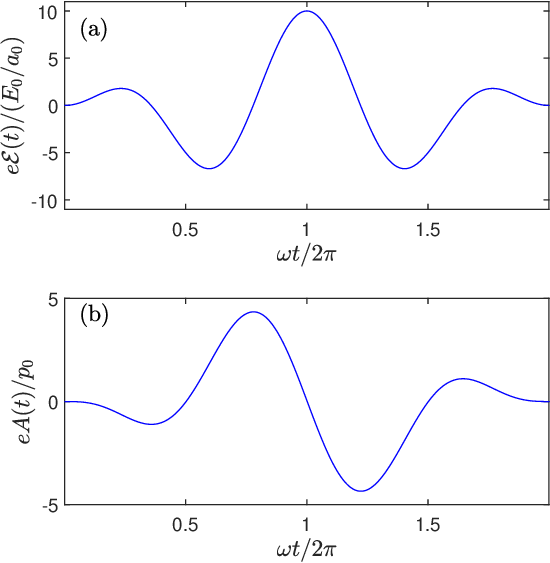}
\caption{The electric field [panel (a)] and the vector potential [panel (b)] plotted at the origin of coordinate system for the model specified by 
Eq.~\eqref{nd3}, and for the laser field parameters: $\omega=2E_0$, $|eA_0|=5p_0$, $N_{\rm rep}=1$, and $N_{\rm osc}=2$.
}
\label{Laser00}
\end{figure}
\begin{figure}[t]
\includegraphics[width=7cm]{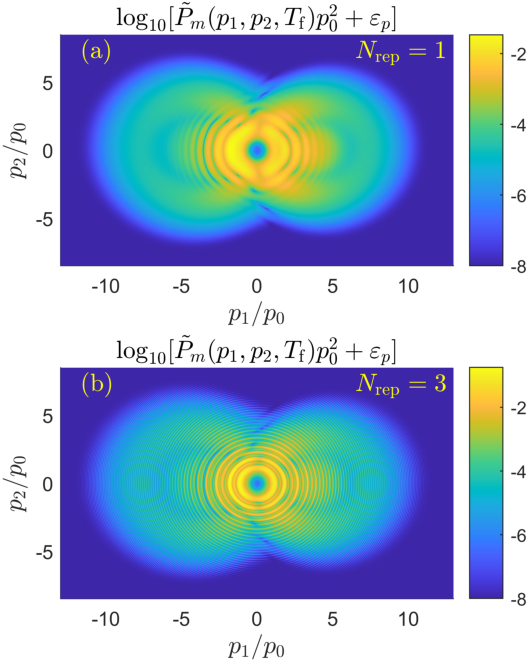}
\caption{Photoelectron momentum distributions~\eqref{nd6} in the Cartesian coordinates for the laser pulse represented in Fig.~\ref{Laser00} [panel (a)]
and for the train comprising of three such pulses [panel (b)]. The distributions are presented in the logarithmic scale, where the values smaller than  
$\varepsilon_p= 10^{-8}$ are eliminated.
}
\label{MomentumDistr1C2rep13a5K}
\end{figure}

We consider a two-dimensional hydrogen atom in the presence of a train of identical XUV pulses. Its minimal coupling Hamiltonian in the position
representation takes the form,
\begin{equation}
\op{H}(t)=\frac{1}{2\me}\bigl(-\ii {\bm \nabla}-e\bm{A}(\bm{x},t)\bigr)^2+V(\bm{x}),
\label{ps4}
\end{equation}
where $V(\bm{x})$ is the potential energy of electron-proton interaction, whereas
\begin{equation}
\bm{A}(\bm{x},t)=(A(t-x_2/c),0)
\label{ps5}
\end{equation}
is the vector potential of the laser field. More specifically, we model the atomic potential as~\cite{Forre}
\begin{equation}
V(\bm{x})=-\mathcal{Z}(x)\alpha c\frac{\mathrm{erf}(x/a_V)}{x},
\label{ps1}
\end{equation}
where $\bm{x}=(x_1,x_2)$ is the two-dimensional position vector with the norm $x=|\bm{x}|=\sqrt{x_1^2+x_2^2}$. Here, we introduce the effective atomic 
number $\mathcal{Z}(x)=1-\lambda_V\exp(-x^2/b^2_V)$. Moreover, the error function $\mathrm{erf}(\cdot)$ softens the singularity at $x=0$ and it 
guarantees that $V({\bm x})$ has the Coulomb tail $-1/x$ for large $x$. With the parameters $a_V=0.1a_0$, $b_V=10a_0$, and $\lambda_V=0.46$
we make sure that the energy of the ground state $E_B$ supported by $V({\bm x})$ is close to the ground state energy of the three-dimensional 
hydrogen atom, $-0.5E_0$. Note that the corresponding bound state wave function, $\psi_B(\bm{x})$, is calculated numerically using the 
Feynman-Kac method for imaginary times. Such a model of hydrogen atom is exposed to the laser field which, according to Eq.~\eqref{ps5}, is linearly
polarized along the $x_1$-axis and it propagates in the $x_2$-direction. Hence, it depends on the retarded time, $t_{\rm r}=t-x_2/c$. 
Moreover, we represent it by the following shape function,
\begin{equation}
A(t_{\rm r})=\begin{cases}
A_0\sin^2\left(\frac{\omega t_{\rm r}}{2N_{\rm osc}}\right)\sin(\omega t_{\rm r}); & 0\leqslant t_{\rm r}\leqslant \frac{2\pi}{\omega}N_{\rm osc}N_{\rm rep}, \cr
0; & \;{\rm otherwise},
\end{cases}
\label{nd3}
\end{equation}
which consists of $N_{\rm rep}$ identical pulses. Each pulse from the train is characterized by its strength $A_0$, carrier wave frequency $\omega$, and
the number of cycles $N_{\rm osc}$. This means that the individual pulse lasts for $\tau_p=2\pi N_{\rm osc}/\omega$. As 
we consider here no time delay between pulses, the train
takes $T_p=N_{\rm rep}\tau_p$. As an example, in Fig.~\ref{Laser00} we demonstrate the single XUV pulse assuming that $|eA_0|=5p_0$, 
$\omega=2E_0$, and $N_{\rm osc}=2$. While in the lower panel we plot Eq.~\eqref{nd3} for $N_{\rm rep}=1$ at the origin of coordinate system 
(${\bm x}={\bm 0}$), in the upper panel we show the corresponding electric field. As $x_2=0$, in Fig.~\ref{Laser00} we write explicitly dependence on $t$. 
More generally, the electric field is defined as $\bm{\mathcal{E}}(t_{\rm r})=-\dot{A}(t_{\rm r}){\bm e}_1={\cal E}(t_{\rm r}){\bm e}_1$, where the dot means 
the time derivative. Let us note that such isolated pulse can be synthesized out of two monochromatic copropagating plane waves of frequencies $\omega$
and $2\omega$. This means that in order to synthesize the corresponding pulse train a similar configuration, consisting of two long pulses with carrier wave 
frequencies $\omega$ and $2\omega$, can be used.

Having specified the model, let us turn to the numerical method of calculations. The calculations are done on a two-dimensional uniform spatial 
grid in Cartesian coordinates. Then, the wave function describing the electron dynamics is calculated 
at every time step of its evolution, which is governed by the Hamiltonian~\eqref{ps4}. As explained in detail in Ref.~\cite{Suster}, this is done
using the Suzuki-Trotter scheme with the split-step Fourier approach. In this way we propagate the initial bound state of the electron, $\psi_B({\bm x})$,
under the presence of the laser field. The evolution time $T_{\rm f}$ is adjusted such that the final electron state is laser-field free,
meaning that $T_{\rm f}>2\pi N_{\rm osc}N_{\rm rep}/\omega$. In order to eliminate from the final electron state $\psi_B({\bm x},T_{\rm f})$ the contributions 
of the bound and low-energy scattering states, we use the mask function (for more details, see Ref.~\cite{KK2025}). As a result, we end up with 
the truncated electron wave function, $\psi_m({\bm x},T_{\rm f})$. Its Fourier transform, $\tilde{\psi}_m({\bm x},T_{\rm f})$, defines the momentum distribution of photoelectrons,
\begin{equation}
\tilde{P}_m({\bm p},T_\mathrm{f})=|\tilde{\psi}_m(\bm{p},T_\mathrm{f})|^2,
\label{nd6}
\end{equation}
which will be presented in either the Cartesian or in the polar coordinates. In our illustrations, we will distinguish the two by writing explicitly 
different sets of arguments, $\tilde{P}_m(p_1,p_2,T_{\rm f})$ or $\tilde{P}_m(E_{\bm p},\varphi_{\bm p},T_{\rm f})$, with the following relations 
$p_1=|{\bm p}|\cos\varphi_{\bm p}$, $p_2=|{\bm p}|\sin\varphi_{\bm p}$, and $|{\bm p}|=\sqrt{2m_{\rm e}E_{\bm p}}$. Note, however, that
$\tilde{P}_m(p_1,p_2,T_\mathrm{f})=\tilde{P}_m(E_{\bm p},\varphi_{\bm{p}},T_\mathrm{f})$.

\begin{figure}[t]
\includegraphics[width=7cm]{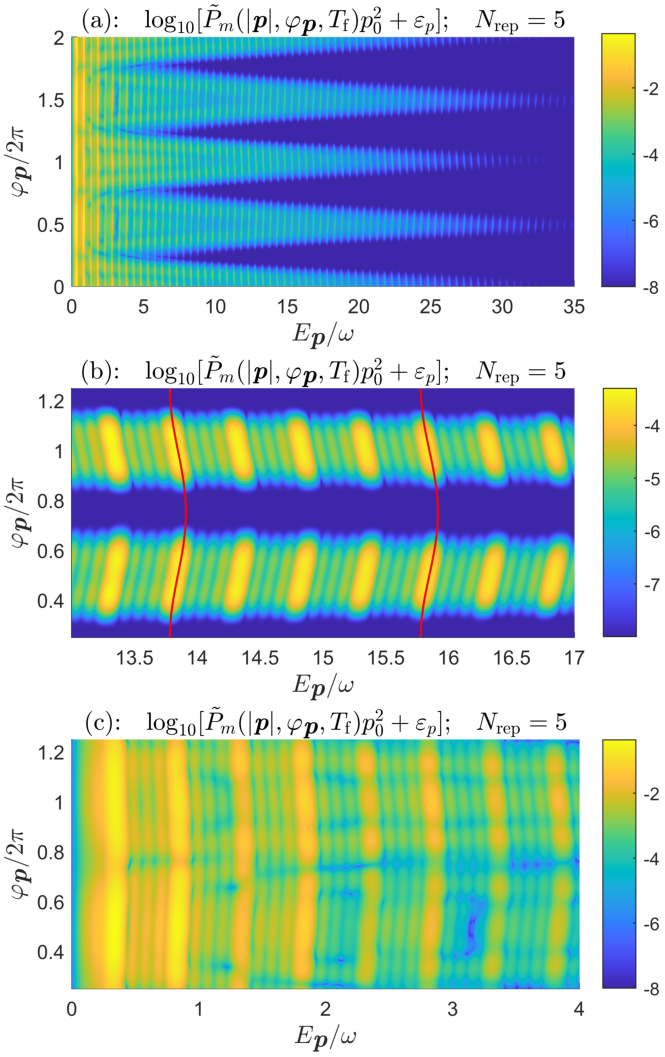}
\caption{Photoelectron momentum distributions~\eqref{nd6} in polar coordinates for the sequence of five XUV pulses ($N_{\rm rep}=5$) shown in Fig.~\ref{Laser00}.
Panel (a) demonstrates the entire energy distribution. On the other hand, panel (b) represents its mid-energy whereas panel (c) shows its low-energy portions.
The distributions are presented in the logarithmic scale, where the values smaller than $\varepsilon_p= 10^{-8}$ are eliminated. The red oscillatory curve
in panel (b) indicates the analytical prediction of the major peak maxima, Eq.~\eqref{fit-red}, which for the current parameters fits nicely with 
the numerical results.
}
\label{Interp800x800XY2RPhi1Cchi000rep5a5x2LK}
\end{figure}

\subsection{Color mappings of photoelectron momentum distributions}

In Fig.~\ref{MomentumDistr1C2rep13a5K}, we demonstrate the momentum distributions of photoelectrons ionized by a single XUV pulse ($N_{\rm rep}=1$) or 
by a train comprising of three such pulses ($N_{\rm rep}=3$) for the model shown in Fig.~\ref{Laser00}. In the case of the single pulse [panel (a)], the distribution
consists of wide rings, which are centered at the proton location. Similar rings, but finer, are observed for the train of pulses in panel (b). In this case,
the rings are interchangeably larger or smaller in magnitude. Namely, in between two subsequent high-intensity rings, there is a low-intensity ring as well.
Such behavior is described by the Fraunhoffer formula~\eqref{abbrev10} that has been derived in Sec.~\ref{sec::sfa} using the QRSFA analysis. According 
to this formula, for a train of $N_{\rm rep}$ pulses, in between subsequent major maxima in the photoelectron momentum or energy spectrum, there
should be also $(N_{\rm rep}-2)$ secondary maxima. This agrees with the results presented in Fig.~\ref{MomentumDistr1C2rep13a5K}. We will come 
back to this aspect later, when analyzing the energy distributions of photoelectrons in the given direction.

\begin{figure}[t!]
\includegraphics[width=7cm]{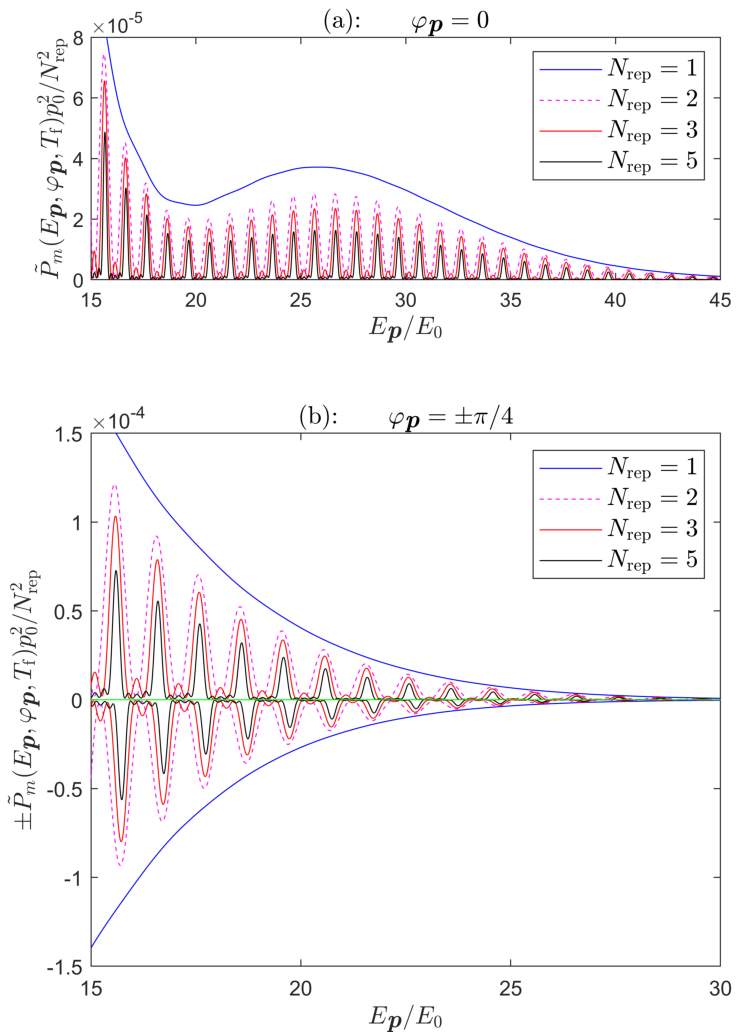}
\caption{Energy distributions of photoelectrons emitted either in the direction specified by the polar angle $\varphi_{\bm p}=0$ [panel (a)] 
or $\varphi_{\bm p}=\pm\pi/4$ [panel (b)]. In panel (b), we plot $\tilde{P}_m(E_{\bm p},\varphi_{\bm p}=\pi/4,T_{\rm f})$ in the upper frame and 
$-\tilde{P}_m(E_{\bm p},\varphi_{\bm p}=-\pi/4,T_{\rm f})$ in the lower frame. Both frames are separated by the solid green line.
Moreover, the presented spectra concern ionization driven by either 
a single XUV pulse (solid blue line) or a train comprising of two (dashed pink line), three (solid red line), and five (solid black line) identical
XUV pulses, each of them being represented in Fig.~\ref{Laser00}. Here, we plot the mid- to high-energy portions of the spectra.
All spectra are divided by $N_{\rm rep}^2$. 
}
\label{combs1C2a5phi0K}
\end{figure}
\begin{figure}[t!]
\includegraphics[width=7cm]{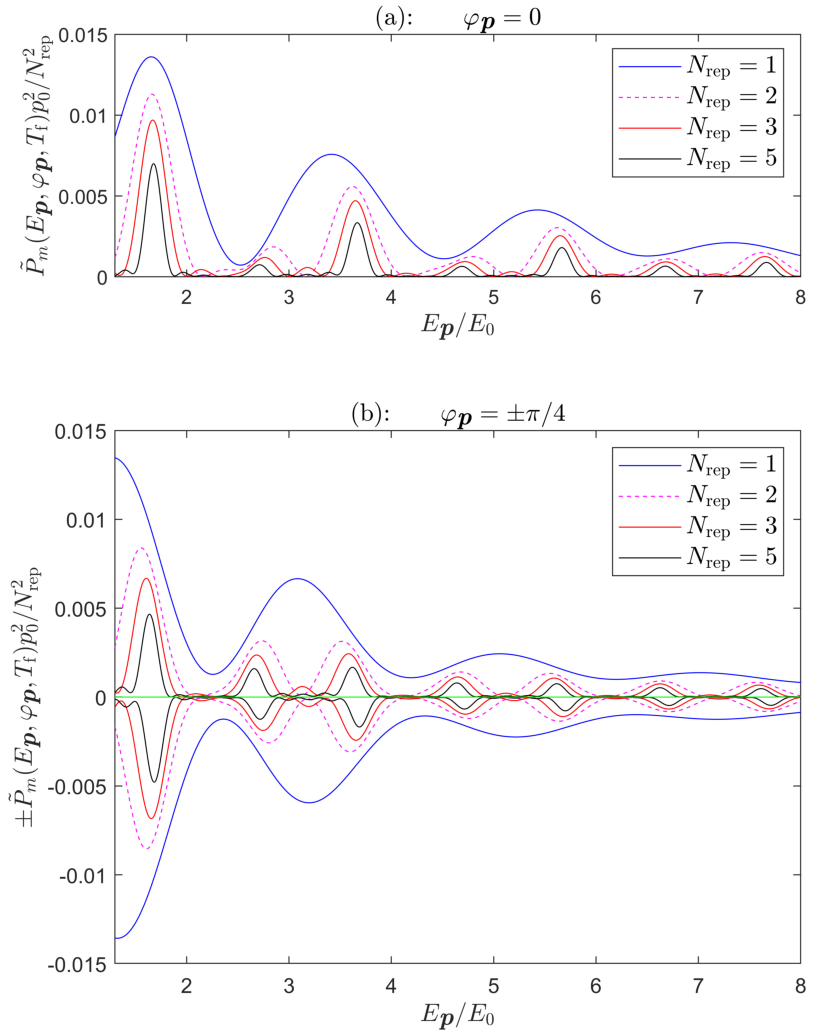}
\caption{Same as in Fig.~\ref{combs1C2a5phi0K} but for the low-energy portions of the distributions.
}
\label{combs1C2a5phi0LK}
\end{figure}

The photoelectron momentum distribution in polar coordinates for even longer pulse train (comprising of $N_{\rm rep}=5$ pulses) is presented in 
Fig.~\ref{Interp800x800XY2RPhi1Cchi000rep5a5x2LK}. While in panel (a) we show the momentum distribution in the entire energy range, its enlarged 
mid-energy portion is presented in panel (b). As one can see, the distribution in panel (a) consists of multiple stripes, which are tilted similar to 
the case when ionization is driven by a long laser pulse considered in Ref.~\cite{KK2025}. In Ref.~\cite{KK2025}, the stripes were separated by the 
photon energy, which allowed us to interpret them as multiphoton peaks. Also, we observed there secondary peaks which were attributed to interference 
caused by the finite character of the laser pulse~\cite{Klaiber}. This time the stripe pattern seems to be of a different origin. First of all, as it follows from
Fig.~\ref{Interp800x800XY2RPhi1Cchi000rep5a5x2LK}(b), the major stripes are separated by roughly $\omega/2$ (or, more generally, by $\omega/N_{\rm osc}$) and between them we observe
three stripes of lower intensity. This behavior matches the Fraunhoffer formula~\eqref{abbrev10}, as already noted in Fig.~\ref{MomentumDistr1C2rep13a5K}(b). 
We conclude, therefore, that the observed patterns originate from the interference of probability amplitudes of ionization by different pulses 
from the train. This has been recognized before as an analog of the Young-type interference phenomenon in the time domain~\cite{Eikonal}.
What has not been observed before in this context is the tilting of the interference fringes. Note that this pattern fits very well with the analytical 
formula derived in Ref.~\cite{KK2025}, 
\begin{equation}
E_{\bm p}(\varphi_{\bm p})=E_{\bm p}-\sqrt{2m_{\rm e}E_{\bm p}}\frac{\langle U_p\rangle}{m_{\rm e}c}\sin\varphi_{\bm p},
\label{fit-red}
\end{equation}
which is represented in Fig.~\ref{Interp800x800XY2RPhi1Cchi000rep5a5x2LK}(b) as the solid red line. $E_{\bm p}(\varphi_{\bm p})$ describes the 
dependence of the energy of the major maximum from the spectrum on the photoelectron emission angle, $\varphi_{\bm p}$, in the case when the interaction 
with the laser field is treated exactly. In contrast, $E_{\bm p}$ in Eq.~\eqref{fit-red} corresponds to the energy of photoelectron emitted along 
the polarization direction of the laser field, i.e., at $\varphi_{\bm p}=0$ or $\pi$. In this case, the laser field does not transfer the momentum
to the electron, and so the given major peak appears at the same energy $E_{\bm p}$ independently of $\varphi_{\bm p}$. Also in 
Eq.~\eqref{fit-red}, we have introduced the time-averaged ponderomotive energy of the electron driven by the laser 
field $\langle U_p\rangle$, which for the current case equals
\begin{equation}
\langle U_p\rangle=\frac{1}{T_p}\int _0^{T_p} \frac{e^2{\bm A}^2(t)}{2m_{\rm e}}\dd t=\frac{3}{16}\frac{e^2A_0^2}{2m_{\rm e}}.
\label{pondero}
\end{equation}
Note that, since all pulses in the train are identical, this quantity does not depend on the train duration. Finally, in Fig.~\ref{Interp800x800XY2RPhi1Cchi000rep5a5x2LK}(c)
we present the low-energy portion of the spectrum from Fig.~\ref{Interp800x800XY2RPhi1Cchi000rep5a5x2LK}(a). A new feature of the spectrum here is that 
the lower the stripe order the smaller amplitude of the oscillations of $E_{\bm p}(\varphi_{\bm p})$. This is in agreement with 
Eq.~\eqref{fit-red}.

\subsection{Comb structures in the photoelectron energy distributions}

While in the last Section we have analyzed color mappings of the photoelectron momentum distributions, here we focus on their energy spectra.
Specifically, in Fig.~\ref{combs1C2a5phi0K}(a) we present the energy spectra of photoelectrons aligned with the XUV field polarization direction,
i.e., for $\varphi_{\bm p}=0$. Here, the solid blue line corresponds to the case when ionization is driven by a single pulse ($N_{\rm rep}=1$) presented
in Fig.~\ref{Laser00}. While in this case the distribution varies slowly with the photoelectron energy, this is changed if $N_{\rm rep}>1$.
The remaining curves represent the data for ionization by a pulse sequence, where the pulse shown in Fig.~\ref{Laser00} is repeated twice (dashed pink line), 
three (solid red line) or five (solid black line) times. Note that all spectra are divided by $N_{\rm rep}^2$. We see that for $N_{\rm rep}>1$ 
the spectra acquire comb-like structures, with the major peaks separated by approximately $\omega/N_{\rm osc}$. Additional peaks also appear for $N_{\rm rep}=3$ and $N_{\rm rep}=5$.
While this is difficult to notice on the scale of the figure, we recognize $(N_{\rm rep}-2)$ secondary peaks and $(N_{\rm rep}-1)$ minima 
between the subsequent major maxima. Those minima resemble zeros of the probability distributions predicted by the Fraunhoffer formula, 
Eq.~\eqref{abbrev10}. The formula predicts also the coherent-like enhancement of the distributions, with the typical scaling of $N_{\rm rep}^2$.
We see from Fig.~\ref{combs1C2a5phi0K}(a), however, a partial loss of coherence as the spectra do not scale like $N_{\rm rep}^2$
for different $N_{\rm rep}$. The same is observed 
for other emission angles, which is illustrated in Fig.~\ref{combs1C2a5phi0K}(b) for $\varphi_{\bm p}=\pm\pi/4$. Both discrepancies follow from the fact 
that Eq.~\eqref{abbrev10} has been derived within the QRSFA approach, which neglects the electron interaction with the residual ion. Once 
the latter is included, the scaled distributions are decreasing in magnitude with increasing $N_{\rm rep}$ and their zeros turn into minima. Note that similar behavior has been 
observed before within the dipole approximation in the framework of generalized eikonal approximation~\cite{Eikonal}.

As mentioned above, in Fig.~\ref{combs1C2a5phi0K}(b) we present the photoelectron energy distributions for either $\varphi_{\bm p}=\pi/4$ (upper frame)
or $\varphi_{\bm p}=-\pi/4$ (lower frame). For visual purposes, in the lower frame we plot the minus corresponding distributions. Both frames are
separated by the solid green line, which marks the zero value of the spectra. While in the dipole approximation, the distributions for 
$\varphi_{\bm p}=\pi/4$ and $-\pi/4$ would peak at the same photoelectron energies, this is not the case here. We observe that for the backward
emission of photoelectrons their spectra are shifted toward larger energies as compared to the case of the forward emission. Thus, it clearly appears
as the nondipole effect, being the direct consequence of Eq.~\eqref{fit-red}.

\begin{figure}[t]
\includegraphics[width=7cm]{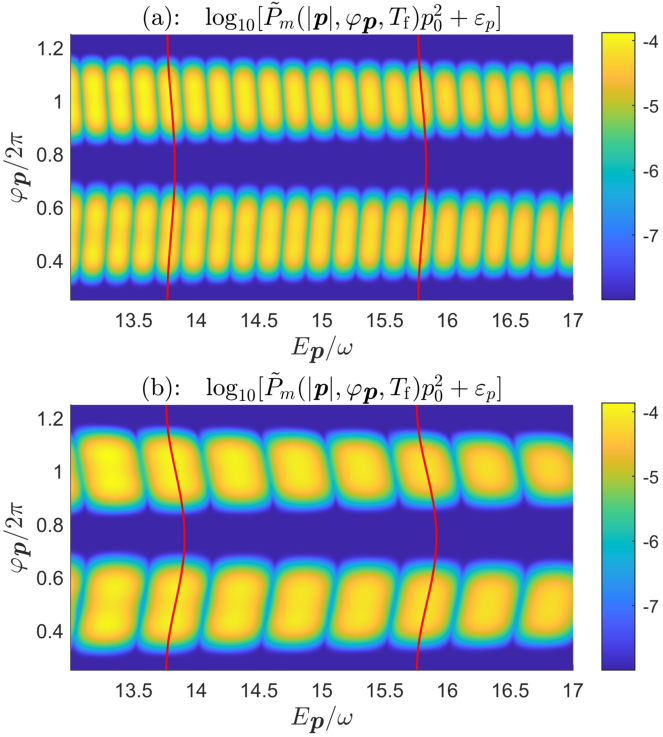}
\caption{Momentum distributions of photoelectrons plotted in the polar coordinates in the case when ionization occurs in a sequence of two  pulses, like the
one shown in Fig.~\ref{Laser00}, either delayed by $\tau_s=6\pi/\omega$ [panel (a)] or $\tau_s=0$ [panel (b)]. The solid wavy lines represent the 
analytical estimate~\eqref{fit-red}. Here, $\varepsilon_p=10^{-8}$ has been introduced for visual purposes.
}
\label{Interp800x800XY2RPhi1Cchi000rep2shiftK}
\end{figure}
\begin{figure}[t]
\includegraphics[width=7cm]{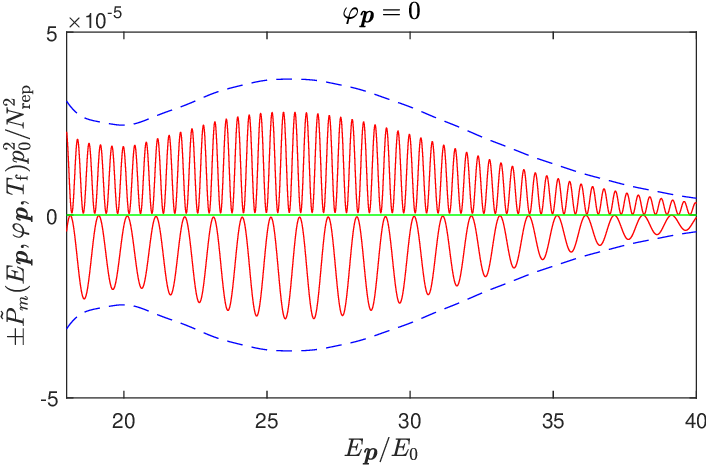}
\caption{Energy spectra of photoelectrons emitted along the polarization direction of the laser field ($\varphi_{\bm p}=0$), that consists of two pulses shown
in Fig.~\ref{Laser00}. While $\tilde{P}_m(E_{\bm p},\varphi_{\bm p},T_{\rm f})$ corresponds to the situation when both pulses are delayed 
in time by $\tau_s=6\pi/\omega$, we plot $-\tilde{P}_m(E_{\bm p},\varphi_{\bm p},T_{\rm f})$ for zero time delay in between them ($\tau_s=0$). Both distributions
are divided by $N_{\rm rep}^2$ and separated by the solid green line. The dashed blue line represents the signal of photoelectrons ionized by a single pulse from the train, Fig.~\ref{Laser00}.
}
\label{combs1C2a5phi0shift}
\end{figure}

The same as in Fig.~\ref{combs1C2a5phi0K} but for the low-energy portion of the photoelectron spectrum is shown in Fig.~\ref{combs1C2a5phi0LK}.
Only this portion of the spectrum is presented, for which the numerical convergence was achieved, i.e., when $E_{\bm p}>1.3E_0$. Here, we observe 
similar but less regular comb structures as compared to Fig.~\ref{combs1C2a5phi0K}. This is because the distribution representing
the ionization driven by a single pulse (solid blue curve) has various modulations. Then, whenever the major peak falls at the minimum of those
modulations, it gets saturated. We also see the shift of the peaks revealing the forward-backward asymmetry. Note that the shift of the peaks increases
with increasing the photoelectron energy.

In this Section we have analyzed the interplay between the photoelectron comb-like structures and rescattering and nondipole effects. For this purpose, we have focused on ionization driven by a train of pulses with no time delay [see, Eq.~\eqref{nd3}]. 
Next, we shall investigate how the time delay between subsequent pulses from the train affects the resulting distributions.

\section{Photoelectrons combs generated by a pulse train with time delay}
\label{timedelay}

We introduce the time delay $\tau_s$ in between pulses from the train that drives ionization.
More precisely, we repeat the pulse presented in Fig.~\ref{Laser00}
$N_{\rm rep}$ times such that every two subsequent pulses are delayed by time $\tau_s$. This means that the pulse from Fig.~\ref{Laser00} 
is repeated at times being multiples of $\tau_d=\tau_p+\tau_s$. In this case, the pulse sequence lasts for $T_p=N_{\rm rep}(\tau_p+\tau_s)$.
In our numerical illustrations, we consider a sequence of two such pulses ($N_{\rm rep}=2$). In Fig.~\ref{Interp800x800XY2RPhi1Cchi000rep2shiftK}, 
we confront the portions of the resulting photoelectron momentum distributions for different time delays: (a) $\tau_s=6\pi/\omega$ and (b) $\tau_s=0$.
Qualitatively they look very similar, as they consists of fringes which follow the oscillatory behavior predicted by Eq.~\eqref{fit-red}.
This analytical prediction is, in fact, represented in both panels as solid wavy lines. One can see, however, that in 
panel (a) the amplitude of those oscillations is smaller than in panel (b). According to Eq.~\eqref{fit-red}, the amplitude of $E_{\bm p}(\varphi_{\bm p})$ is determined by 
the time-averaged ponderomotive energy of the photoelectron $\langle U_p\rangle$. It is straightforward to calculate that for $\tau_s=6\pi/\omega$,
\begin{equation}
\langle U_p\rangle=\frac{2}{5}\cdot\frac{3}{16}\frac{e^2A_0^2}{2m_{\rm e}},
\label{bs}
\end{equation}
which is $2/5$ times smaller than for $\tau_s=0$ [Eq.~\eqref{pondero}]. This means that the amplitude of oscillations in 
Fig.~\ref{Interp800x800XY2RPhi1Cchi000rep2shiftK}(a) is smaller by a factor of $2/5$ compared to Fig.~\ref{Interp800x800XY2RPhi1Cchi000rep2shiftK}(b).
At the same time, the number of stripes in panel (a) is increased by a factor of 5/2. Note the lack of the secondary maxima, which for a train 
comprising of two pulses is in agreement with the Fraunhoffer formula [Eq.~\eqref{abbrev10}].

\begin{figure}[t]
\includegraphics[width=7cm]{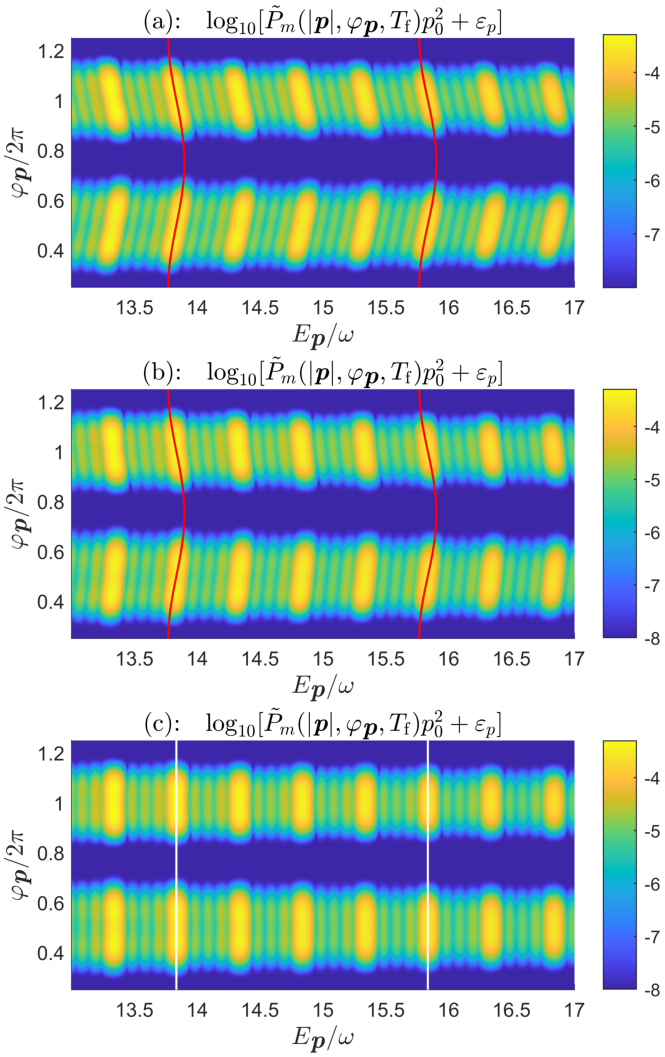}
\caption{Color mappings of the photoelectron momentum distributions calculated by either the strict numerical integration of the Schr\"odinger
equation [panel (a)], in the first-order nondipole approximation [panel (b)], or in the dipole approximation [panel (c)]. We assume that
the ionization occurs in a train of five laser pulses ($N_{\rm rep}=5$) with no time delay ($\tau_s=0$), each of them described by the vector potential 
shape function presented in Fig.~\ref{Laser00}. In the bottom panel, the chosen main maxima are marked with the solid white lines. In the top and middle 
panels, the corresponding dependence $E_{\bm p}(\varphi_{\bm p})$ specified by Eq.~\eqref{fit-red} is plotted as solid red lines. For visual purposes, we have introduced
$\varepsilon_p=10^{-8}$.
}
\label{Interp800x800XY2RPhi1CXchi000rep5a5x1K}
\end{figure}
\begin{figure}[t]
\includegraphics[width=7cm]{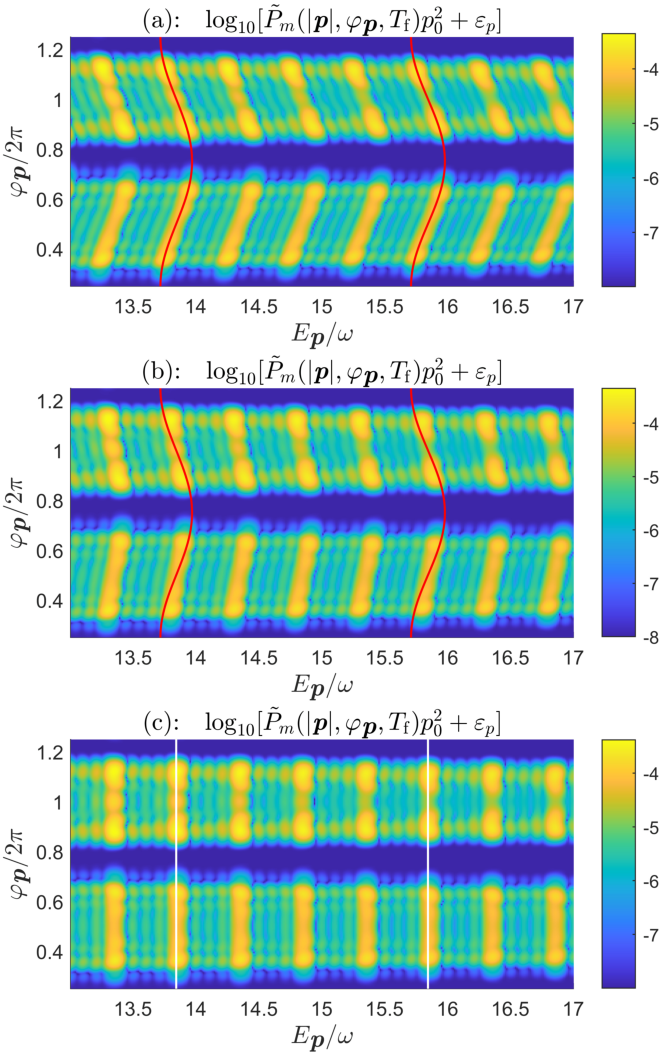}
\caption{Same as in Fig.~\ref{Interp800x800XY2RPhi1CXchi000rep5a5x1K} but for a stronger laser field, with $|eA_0|=7p_0$.
}
\label{Interp1100x800XY2RPhi1CXchi000rep5a7x1K}
\end{figure}

In Fig.~\ref{combs1C2a5phi0shift}, we present the scaled energy spectra of photoelectrons emitted at $\varphi_{\bm p}=0$ by a train of two XUV pulses,
each of them presented in Fig.~\ref{Laser00}. For visual purposes, we
plot $\tilde{P}_m(E_{\bm p},\varphi_{\bm p},T_{\rm f})$ for the case when both pulses are delayed in time by $\tau_s=6\pi/\omega$, whereas
$-\tilde{P}_m(E_{\bm p},\varphi_{\bm p},T_{\rm f})$ is for zero time delay in between them ($\tau_s=0$). Both spectra are separated by the solid green line.
They clearly exhibit the comb structures. In addition, the dashed blue envelopes are marked there as the reference. They describe plus and minus the energy spectrum 
for ionization driven by a single pulse. 
Again, we observe the loss of coherence, as the spectra do not scale like $N_{\rm rep}^2$. Their modulations, however, clearly follow the dashed lines.
One can easily see that the comb structure is roughly 5/2 times denser in the upper frame. Such comparison demonstrates that the structures
originate from the interpulse interference of ionization probability amplitudes and that delay between pulses comprising the train controls their density.

So far, we have studied the appearance of comb structures in ionization using the framework where the interaction with the laser field is treated 
exactly. However, an interesting point to address is how the structures would be affected by typically used approximations such as the dipole or the 
first-order nondipole approximations. This will be analyzed next.

\section{Comparison of the exact TDSE results with different approximations}
\label{comparison}
\begin{figure}[t]
\includegraphics[width=7cm]{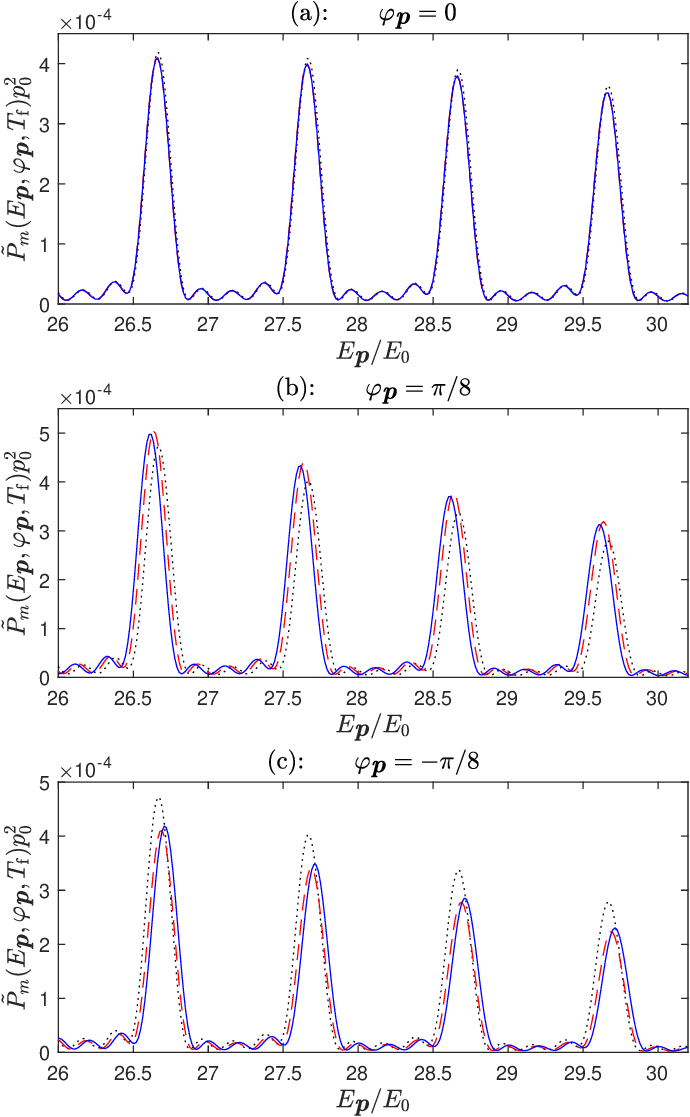}
\caption{Energy distributions of photoelectrons calculated at different angles, as indicated in the figure. The remaining parameters are the same as
in Fig.~\ref{Interp800x800XY2RPhi1CXchi000rep5a5x1K}. Each panel shows comparison between
three approached: the strict TDSE calculation (solid lines), the first-order nondipole approximation (dashed lines), and the dipole approximation 
(black lines). While for $\varphi_{\bm p}=0$ the three lines almost coincide, for $\varphi_{\bm p}=\pi/8$ the spectra calculated rigorously
and in the first-order nondipole approximation are red-shifted whereas for $\varphi_{\bm p}=-\pi/8$ they are blue-shifted as compared to the dipole 
approximation spectra.
}
\label{combscompare1C5chi000a5f0}
\end{figure}

\begin{figure}[t]
\includegraphics[width=7.5cm]{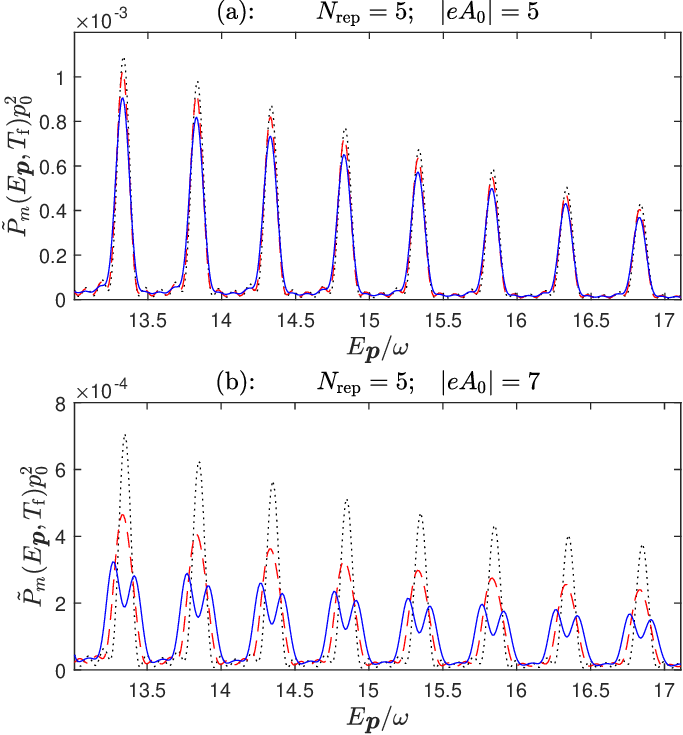}
\caption{Comparison of the angle-integrated energy distributions of photoelectrons calculated based on the strict numerical integration of TDSE
(solid lines), the first-order nondipole approximation (dashed lines), and the dipole approximation (dotted lines). The upper panel
is for $|eA_0|=5p_0$ while the lower panel is for $|eA_0|=7p_0$. The remaining parameters are the same as in 
Fig.~\ref{Interp800x800XY2RPhi1CXchi000rep5a5x1K}.
}
\label{Interp800x800XY2RPhi1CXchi000a57rep5}
\end{figure}

In this Section, we compare the results of three approaches treating the interaction with the laser field: exactly, in the first-order 
nondipole approximation, and in the dipole approximation. In the first case, we strictly solve the time-dependent Schr\"odinger equation 
accounting for the full spatial and temporal dependence of the laser field, as specified by Eqs.~\eqref{ps4} and~\eqref{ps5}. This method has
been used so far in this paper. On contrary, in the first-order nondipole approximation, the vector potential shape function in Eq.~\eqref{ps5} is simplified such that
$A(t-x_2/c)\approx A(t)+x_2{\cal E}(t)/c$, whereas in the dipole approximation we account only for its time dependence, $A(t-x_2/c)\approx A(t)$.
Here, ${\cal E}(t)=-\dot{A}(t)$ represents the shape function of the corresponding electric field.

For numerical illustrations, we consider ionization by a sequence of five laser pulses ($N_{\rm rep}=5$), which are not delayed in time ($\tau_s=0$). 
This guarantees that the comb structures are most spread as compared to other choices of $\tau_s>0$. We start by comparing both approximations with 
the TDSE results presented in 
Fig.~\ref{Interp800x800XY2RPhi1Cchi000rep5a5x2LK}(b). For convenience of the reader, Fig.~\ref{Interp800x800XY2RPhi1Cchi000rep5a5x2LK}(b) is repeated 
in panel (a) of Fig.~\ref{Interp800x800XY2RPhi1CXchi000rep5a5x1K}. Figs.~\ref{Interp800x800XY2RPhi1CXchi000rep5a5x1K}(b) and~\ref{Interp800x800XY2RPhi1CXchi000rep5a5x1K}(c) show the results of
the first-order nondipole approximation and the dipole approximation, respectively. The solid white line in panel~(c) marks the chosen major peaks,
whereas the red oscillating lines in the remaining panels represent their analogues determined by Eq.~\eqref{fit-red}. As it is clear from panel~(c),
in the dipole approximation the comb maxima occur at the same energies, regardless of the photoelectron emission angle $\varphi_{\bm p}$. On contrary,
once the laser field propagation is accounted for, either in the approximate way [panel (b)] or exactly [panel (a)], the comb peaks will appear
at different energies depending on $\varphi_{\bm p}$. The energy shift is not very pronounced for the given laser field parameters, but it increases
with increasing the laser field strength. This is illustrated in Fig.~\ref{Interp1100x800XY2RPhi1CXchi000rep5a7x1K}, which has been calculated
for the same parameters as Fig.~\ref{Interp800x800XY2RPhi1CXchi000rep5a5x1K}, except that now $|eA_0|=7p_0$. In this case, the amplitude of
oscillations is larger compared to Fig.~\ref{Interp800x800XY2RPhi1CXchi000rep5a5x1K}. Also, the discrepancy between the exact results and the results
obtained in the first-order nondipole approximation is more pronounced for a stronger laser field. 

In Fig.~\ref{combscompare1C5chi000a5f0}, we present the energy distributions of photoelectrons emitted at different angles $\varphi_{\bm p}$, which are 
indicated in each panel. The remaining parameters are the same as in Fig.~\ref{Interp800x800XY2RPhi1CXchi000rep5a5x1K}. 
The spectra have been calculated
either in the dipole approximation (dotted lines), in the first-order nondipole approximation (dashed lines), or exactly (solid lines).
Note that for as long as the photoelectrons are ionized along the polarization direction of the laser field [$\varphi_{\bm p}=0$; panel (a)], 
all methods give almost identical results. This follows from the fact that the momentum transfer from the laser field vanishes if photoelectrons 
are emitted at $\varphi_{\bm p}=0$.
For other angles, except of the spectra calculated in the dipole approximation, we observe either blue or red shift of the spectra,
in compliance with Figs.~\ref{Interp800x800XY2RPhi1CXchi000rep5a5x1K} and~\ref{Interp1100x800XY2RPhi1CXchi000rep5a7x1K}. 

Finally, in Fig.~\ref{Interp800x800XY2RPhi1CXchi000a57rep5} we demonstrate the angle-integrated energy distributions,
\begin{equation}
\tilde{P}_m(E_{\bm p},T_{\rm f})=\int_0^{2\pi} d\varphi_{\bm p}\,\tilde{P}_m(E_{\bm p},\varphi_{\bm p},T_{\rm f}),
\label{angle-int}
\end{equation}
when the train of five XUV pulses ($N_{\rm rep}=5$) is driving ionization, for either (a) $|eA_0|=5p_0$ or (b) $|eA_0|=7p_0$. In both panels
we compare the spectra calculated  when the laser field is treated exactly (solid lines), in the first-order nondipole approximation (dashed lines), 
or within the dipole approximation (dotted lines). For the weaker laser field all spectra look qualitatively similar. For the stronger  field, however,
the energy distribution arising from the exact treatment of the laser field looks essentially different, 
as each peak exhibits a double-hump structure. This is not the case for spectra calculated within either approximation. Nor for the experimental data
published in~\cite{Worner2024}. This also suggests that the typically used approximations work reasonably well for weaker fields. They fail, however, when 
the XUV field is strong enough, as presented above for $|eA_0|=7p_0$.

\section{Conclusions}
\label{conclusions}

We have studied nondipole effects in ionization driven by a sequence of identical XUV pulses. This was based on numerical solution of the 
time-dependent Schr\"odinger equation treating the interaction with the laser field exactly. As such calculations are very demanding, at least 
for the considered set of parameters, we had to limit our model to two dimensions. It was demonstrated that, under given circumstances, the comb
structures in the momentum and energy distributions are observed. Their detailed structure, however, is significantly affected by the long-range atomic
interaction and nondipole effects.

First of all, it is expected that the comb structures are coherently enhanced with increasing the number of pulses comprising the train, $N_{\rm rep}$. This follows 
from the Fraunhoffer formula, which has been derived in this paper using the quasi-relativistic SFA. Note that in SFA, the long-range atomic interaction 
is disregarded in the final electronic state. Once it is accounted for, as it is in our numerical calculations, the qualitative features of the comb 
structures are essentially preserved. Except that now the combs do not scale like $N_{\rm rep}^2$ and do not necessarily take zero values. Such loss of coherence can be attributed to 
rescattering, which is inherently accounted for in our numerical analysis. 

The nondipole effects, on the other hand, clearly manifest themselves in our rigorous TDSE momentum and energy distributions of photoelectrons if confronted against 
the dipole approximation. While for the latter, the fringes maxima in photoelectron momentum distributions do not depend on the electron emission angle $\varphi_{\bm p}$, 
this is not the case going beyond the dipole approximation. Specifically, we have observed that the positions of the maxima
start to oscillate with $\varphi_{\bm p}$. As a result, the peak structure observed in the energy distributions is either blue- or red-shifted
as compared to the dipole approximation, depending on the photoelectron emission angle.

In fact, we have compared our rigorous TDSE results with the first-order nondipole approximation as well. Already for smaller laser fields, there appear
small differences in both approaches. Specifically, the tilting of the oscillations of maxima in the momentum distributions is 
already quite different for small laser fields. It becomes more pronounced, however, with increasing the field strength. This is also visible in the 
angle-integrated energy distributions of photoelectrons. Our rigorous TDSE calculations result in the double-hump structure of the comb peaks,
which is not the case when the laser field is treated approximately for the considered intensities.

Finally, we have checked how the time delay between the laser pulses driving ionization may affect the comb structures. We have seen that the more 
separated in time the pulses are, the more dense combs become. This confirms that the comb structures originate from the interference between 
the probability amplitudes of ionization by different pulses or, in other words, from interpulse interferences.

\section*{Acknowledgements}
This work was supported by the National Science Centre (Poland) under Grant No. 2018/30/Q/ST2/00236.

\end{document}